\documentstyle[twoside,fleqn,espcrc2,epsf]{article}
\def\fig#1#2#3{\epsfxsize=#3truein
\vskip -0.3 truein
\centerline{\epsffile{fig#1.eps}}
\centerline{\vbox{{\bf \noindent Figure #1.} #2}}
\bigskip}

\def\figsize{2.2}

\def\one{$S_{eff}$ in topological sector 0}
\def\two{$S_{eff}$ in topological sector 1}
\def\three{Time history of $W$ at $m_f = 0.01$}
\def\four{$\langle W \rangle$ vs. $m_f$}
\newcommand{\AmS}{{\protect\the\textfont2
  A\kern-.1667em\lower.5ex\hbox{M}\kern-.125emS}}

\title{
\vspace{-5.0cm}
\begin{flushright}
{\normalsize CU--TP--774}\\
\vspace*{3.5cm}
\end{flushright}
Domain Wall Fermions and MC Simulations of Vector Theories.}

\author{P. Vranas\address{Dept. of Physics, Columbia University,
	New York, NY 10027}}
       
\begin{document}

\begin{abstract}

It is known that domain wall fermions may be used in MC simulations of
vector theories. The practicality and usefulness of such an
implementation is investigated in the context of the vector Schwinger
model, on a 2+1 dimensional lattice. Preliminary results of a Hybrid
Monte Carlo simulation are presented.

\end{abstract}

\maketitle

\section{Introduction}

Domain Wall Fermions (DWF) for lattice gauge theories were developed
in \cite{DK}.  Since then a wealth of activity has followed with the
main focus being the application of DWF to chiral lattice gauge
theories, but also to vector lattice gauge theories (see \cite{LP} and
references therein). As a result of these works little doubt remains
that DWF can be used to regularize vector gauge theories on the
lattice.  Because DWF have many attractive features it is surprising
that little work has been done in the past few years investigating
their practicality and usefulness in numerical simulations of lattice
vector gauge theories. A preliminary investigation of this, in the
context of the vector Schwinger model, is presented here.

\section{The model}

The massless case of the one flavor model is described by the lattice
Euclidean action:
\begin{eqnarray}
S[\bar\psi,\psi,A_\mu] = \beta [ 1 - \sum_p Re(U_p)] 
- \bar\Psi {\cal M}(U) \Psi 
\nonumber
\end{eqnarray}
\vskip -0.4 truecm
\begin{eqnarray}
{\cal M}(U)_{(n;s),(n^\prime;s^\prime)} = \delta_{s,s^\prime} \sum_{\mu = 1}^2 \big[ 
{(1+\gamma_\mu)\over 2} U_{n, \mu} \delta_{n+\hat\mu, n^\prime}  
\nonumber
\end{eqnarray}
\vskip -0.5 truecm
\begin{eqnarray}
+ {(1-\gamma_\mu)\over 2} U^{\dagger}_{n-\hat\mu, \mu} 
\delta_{n-\hat\mu, n^\prime} \big] 
\nonumber
\end{eqnarray}
\vskip -0.5 truecm
\begin{eqnarray}
+ \delta_{n,n^\prime} \big[ {(1+\gamma_5)\over 2} 
\delta_{s+1,s^\prime} + {(1-\gamma_5)\over 2} 
\delta_{s-1,s^\prime}\big]
\nonumber
\end{eqnarray}
\vskip -0.5 truecm
\begin{eqnarray}
- \delta_{n,n^\prime} \delta_{s,s^\prime} [3 - m(s)]
\label{action}
\end{eqnarray}

In the above $U$ is the U(1) gauge field, $U_p$ the standard
plaquette, $\beta = g_0^{-2}$ with $g_0$ the gauge coupling, and $n$
is a collective space time coordinate. The photon mass is $m_\gamma =
{g_0 / \sqrt{\pi}}$ and the space time volume is $V=L^2$.  The key
ingredient of DWF is the introduction of the extra direction $s$, with
periodic boundary conditions, size $L_s$, and a mass defect at $s=0$
and $s=L_s/2$ ($m(s) = +m0$ for $0\leq s < L_s/2$ and $m(s) = -m0$ for
$L_s/2\leq s < L_s, 0< m_0 < 1$).  In the free case a Dirac fermion
appears with its right component exponentially peaked at $s=0$ and its
left one at $s=L_s/2$.  The overlap of the left and right components
is small and decreases exponentially with increasing $L_s$.  In the
limit $L_s \rightarrow
\infty$ one obtains a single massless Dirac fermion.  The introduction
of the extra direction also introduces heavy modes.  Their
contribution needs to be subtracted \cite{NN1}. This is done by
dividing the fermionic determinant, $\det[{\cal M}]$ by
$\sqrt{\det[{\cal M}_+] \det[{\cal M}_-]}$, where ${\cal M}_+$, ${\cal
M}_-$ are the same as ${\cal M}$ but with $m(s) = +m_0$, $m(s) = -m_0$
respectively. These determinants can be produced by introducing
appropriate auxiliary bosonic fields.

There are two implementations of DWF that can be used in numerical
simulations:

\noindent
{\bf I} The overlap formalism \cite{NN1}. A transfer matrix is
constructed along the extra direction and from it the corresponding
Hamiltonian $\cal H$ is extracted.  This formulation allows for the strict $L_s
\rightarrow \infty$ limit to be taken.  Observables can be calculated
by obtaining all the eigenvalues and eigenvectors of $\cal H$ ($\cal
H$ is a matrix of size $\sim V \times V$).  The lattice vector
Schwinger model was simulated successfully using this method
\cite{NNV}. However, the method requires large amounts of computer time and high
statistics simulations in four dimensions and large volumes
may not be possible at present.

\noindent
{\bf II} A direct simulation of (\ref{action}) with $L_s$ finite. The
chiral limit is obtained by extrapolating to the $L_s \rightarrow
\infty$ limit. This method is very attractive for several reasons:

\noindent
{\bf a)} Unlike Wilson fermions there is no fine tuning involved; the
chiral limit is the $L_s \rightarrow \infty$ limit.  At the $L_s
\rightarrow \infty$ limit the quark mass is multiplicatively
renormalized \cite{FS} (for a recent investigation of this using
Hybrid Monte Carlo (HMC) see \cite{AJ}). Although for finite $L_s$ there
will always be an additive component its size is expected to be
decreasing exponentially with $L_s$.

\noindent
{\bf b)} Unlike staggered fermions there is no breaking of flavor
symmetry on the lattice.

\noindent
{\bf c)} Because the overlap of the left and right components is
exponentially small in the free case, one would expect that very small
explicit breaking of chiral symmetry can be obtained for a relatively
small $L_s$. If this is the case, then the method will be very valuable
in studies involving spontaneous chiral symmetry breaking.

\noindent
{\bf d)} Anomalous symmetry breaking has a natural interpretation
\cite{DK} and one may expect that this will facilitate related
studies.

Of course, there is a price to be paid for these nice features. The
theory has one more dimension and therefore demands more
computer resources.

Given the above considerations several questions should be answered
before DWF can be used to simulate lattice vector gauge theories and
in particular lattice QCD. Are there any hidden difficulties in an HMC
simulation of the model? Does the method work and give the correct
answers?  How does the computational difficulty depend on $L_s$ and
how large should $L_s$ be? How does the computational difficulty of
DWF compare with Wilson or Staggered fermions?  How good are DWF in
addressing questions related to anomalous breaking of axial
symmetries and spontaneous breaking of chiral symmetries?  Preliminary
answers to a few of these questions are presented below.

\section{The size of the extra dimension}

The $L_s$ dependence can be investigated by direct comparison with the
$L_s \rightarrow \infty$ result obtained using the overlap formalism.
In the massless case the fermionic effective action $S_{eff}$ is finite 
in the zero topological sector and $S_{eff}
\rightarrow -\infty$ otherwise. On an $8 \times 8$ lattice, $S_{eff}$
is calculated for several $L_s$ for a background gauge field with
topological charge $0$ in figure 1, and with charge $1$ in figure 2. The very
different $L_s$ dependence in the two topological sectors is evident.
This indicates that DWF correctly reproduce topological effects
already at $L_s= 10 - 12$. Furthermore, the overlap result in the zero
topological sector (dashed line in figure 1) is also reached at $L_s=
10 - 12$. Similar results for the size of $L_s$ have been obtained
from studies of the pion mass in the $0$ sector \cite{AJ}.
\fig{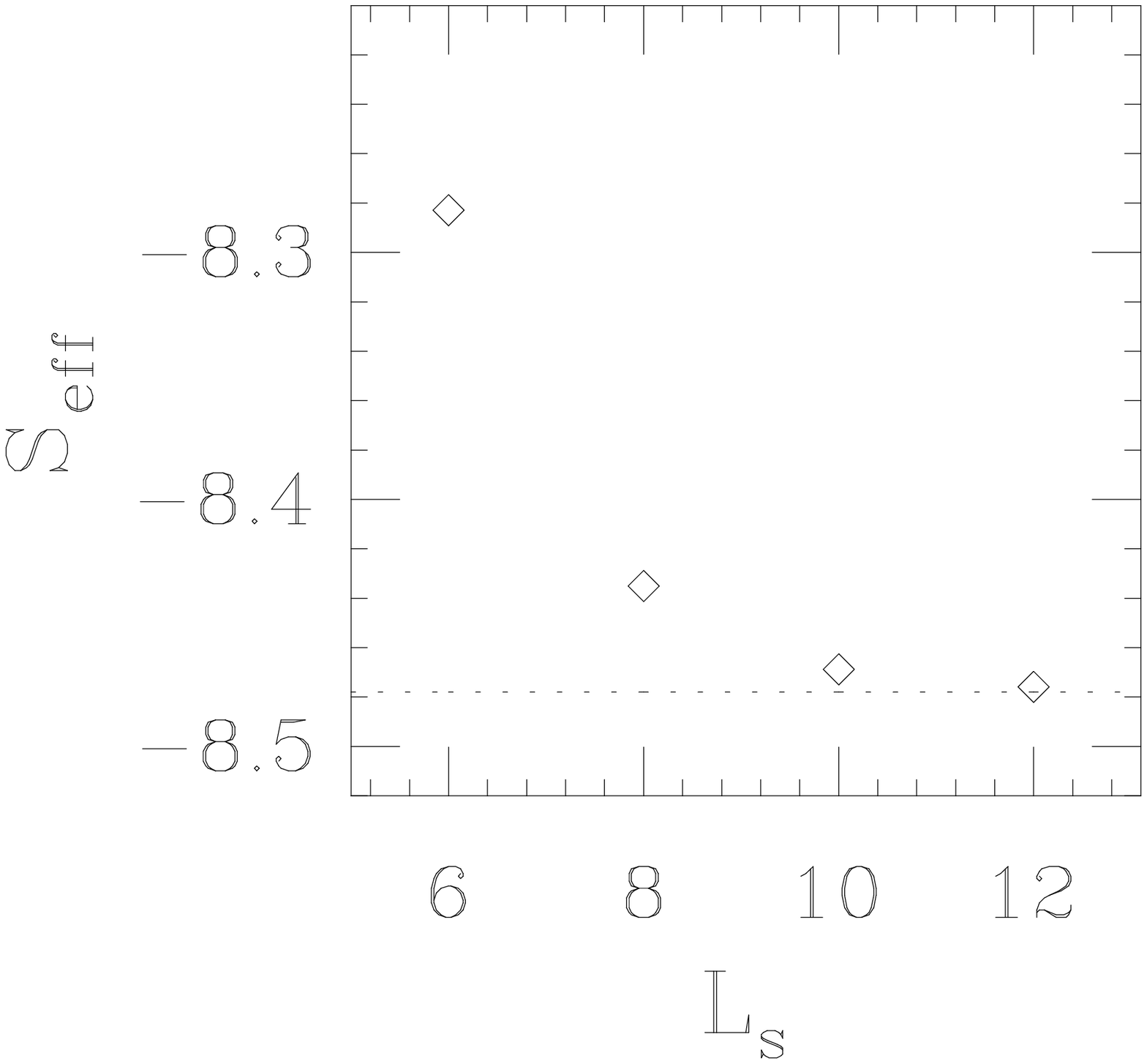}{\one}{\figsize}
\fig{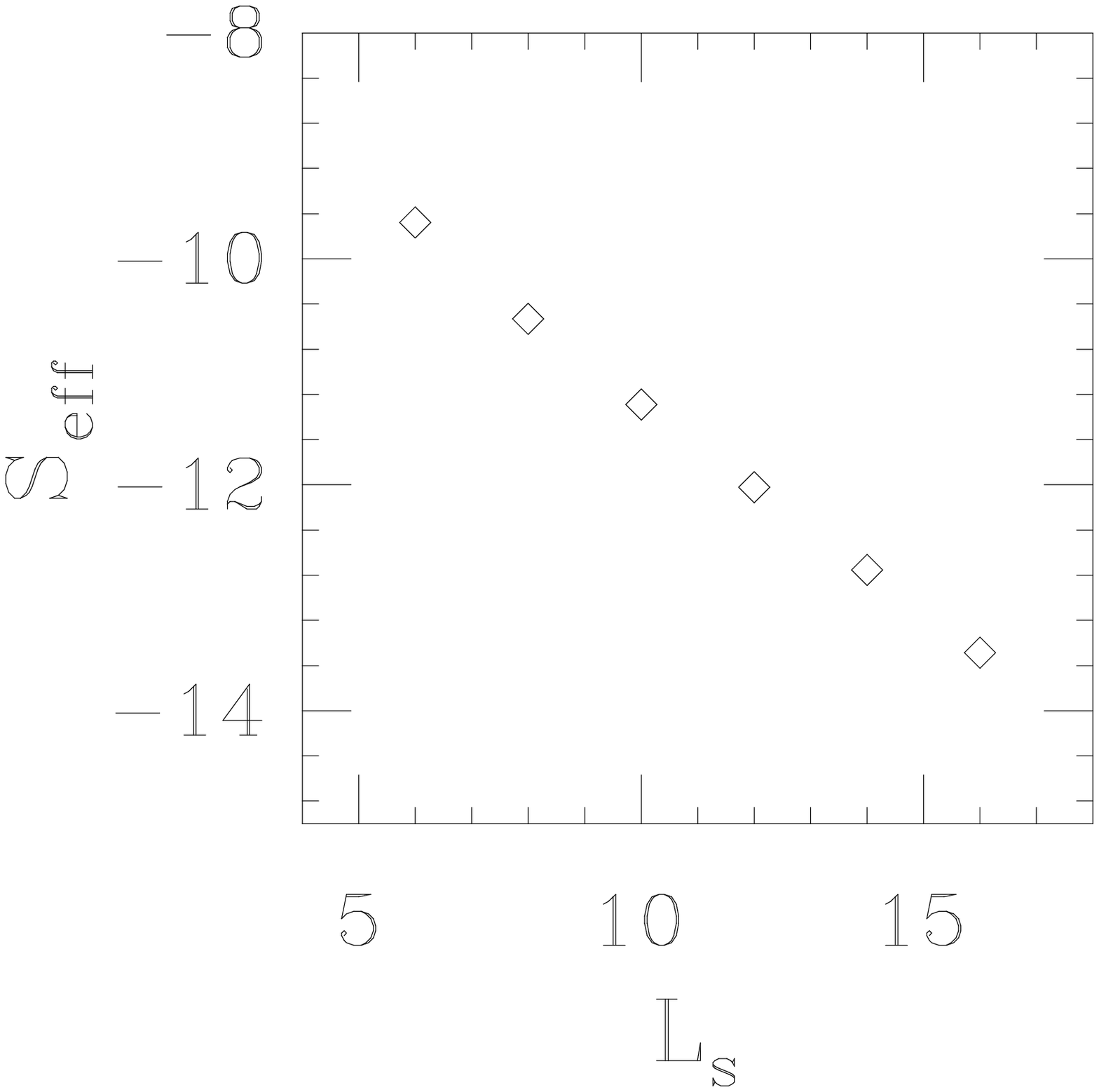}{\two}{\figsize}

\section{Hybrid Monte Carlo and anomalous symmetry breaking}

A variant of (\ref{action}) developed in \cite{FS} is most suited for
use with the HMC algorithm. The basic difference is that $m(s)=m_0$ but
free boundary conditions are used along the extra direction. The two
chiral modes are now bound to the two free boundaries. A mass term
connects the free boundaries and introduces a bare quark mass $m_f$.
The advantage of this is that it only needs half the size
of $L_s$ and also only a single bosonic field to perform the
subtraction mentioned earlier.

The massive $N_f = 2$ flavor vector Schwinger model is simulated using
this formulation and the HMC algorithm. The expectation value of the
operator
$
W = \prod_{i=1}^{N_f} \bar\Psi^i_R \Psi^i_L + \prod_{i=1}^{N_f}
\bar\Psi^i_L \Psi^i_R
$
is calculated.
If the volume is kept fixed and $m_f$ is made very small
the effect of the zero modes coming from different topological sectors
becomes important \cite{LS}.  In particular, as can be seen from 
the overlap implementation \cite{NN1} \cite{NNV}, 
in the massless and $L_s \rightarrow \infty$ limit the
operator $W$ receives contributions only from sectors $\pm 1$, while
the fermion Boltzman weight (fermion determinant) is not zero only in
sector 0.  As $m_f$ is turned on (and/or $L_s$ is decreased from
infinity) the fermionic determinant becomes non zero in sectors
other than $0$ and the operator $W$ receives contributions from
sectors other than $\pm1$.  Therefore for small $m_f$ the HMC
algorithm will mostly sample the sector $0$ where the observable $W$
receives small contributions. The algorithm will infrequently
visit the sectors $\pm 1$ but when it does the observable $W$ will
receive large contributions to make up for the small sampling
rate. As a result, when $m_f$ is decreased a larger number of HMC
iterations will be needed to sample the $\pm 1$ sectors correctly.

\section{Preliminary results}

Preliminary results are presented for $m_0=0.9$, $L=6 $, $m_\gamma
L=3.0$.  The HMC sweeps vary from $6,000$ for $m_f=0.01$ to $2,000$
for $m_f=0.5$. The trajectory length is $1$ and the step size is
$0.02$ for $m_f=0.01$ and $0.04$ for the rest. The average conjugate
gradient iterations vary from $56$ to $86$. The time history of $W$
for $m_f=0.01$ is shown in fig. 3. The large ``spikes'' are related to
configurations with charge $\pm 1$ (measured with the geometric
method).  $\langle W\rangle$ vs. $m_f$ is in fig. 4 and agrees
roughly with the $m_f=0$ overlap result \cite{NNV}.  For comparison
with the exact answer see \cite{NNV}.

This work was supported by DOE grant \# DE-FG02-92ER40699.

\fig{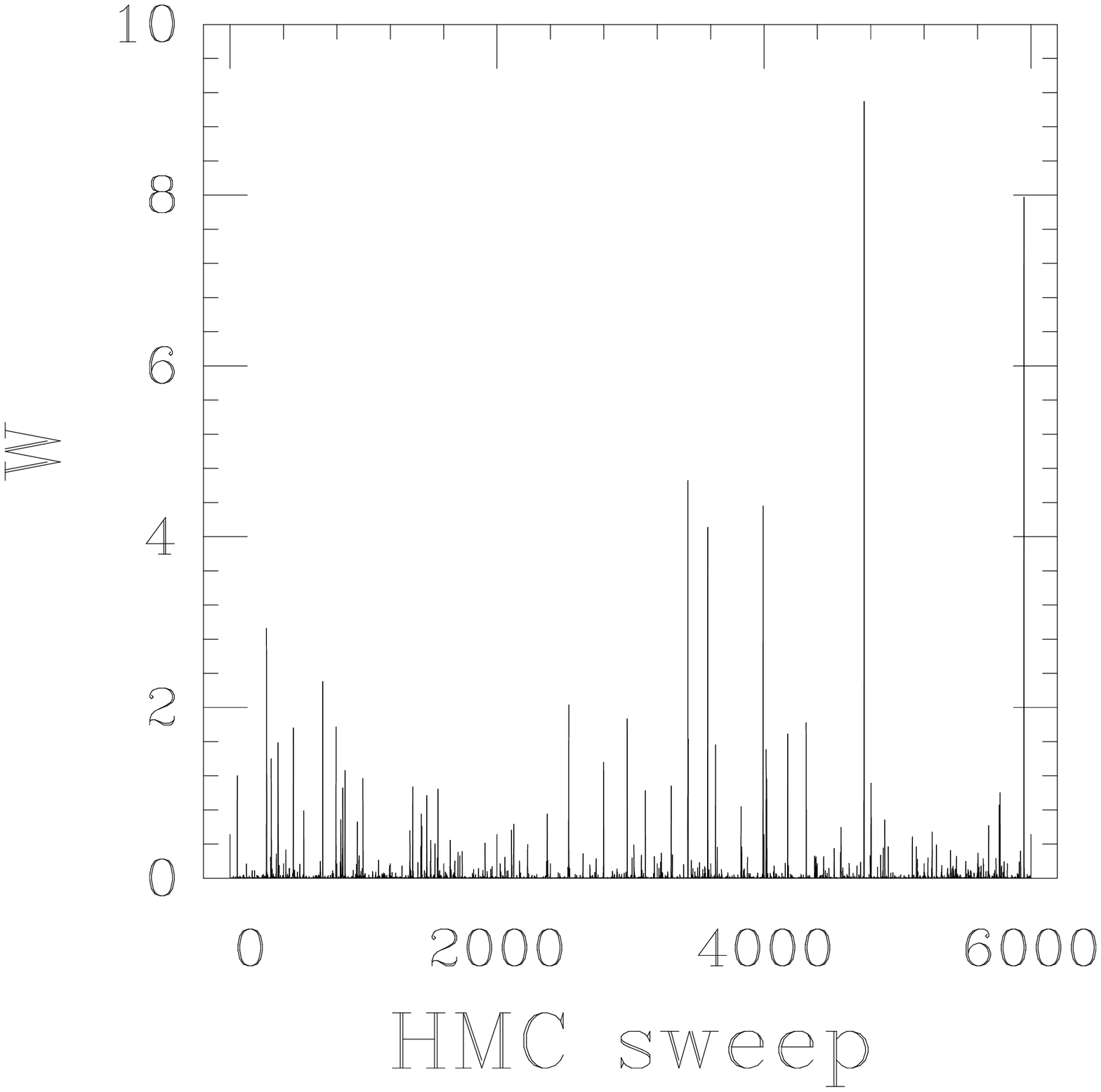}{\three}{\figsize}

\vskip 0.3 truecm

\fig{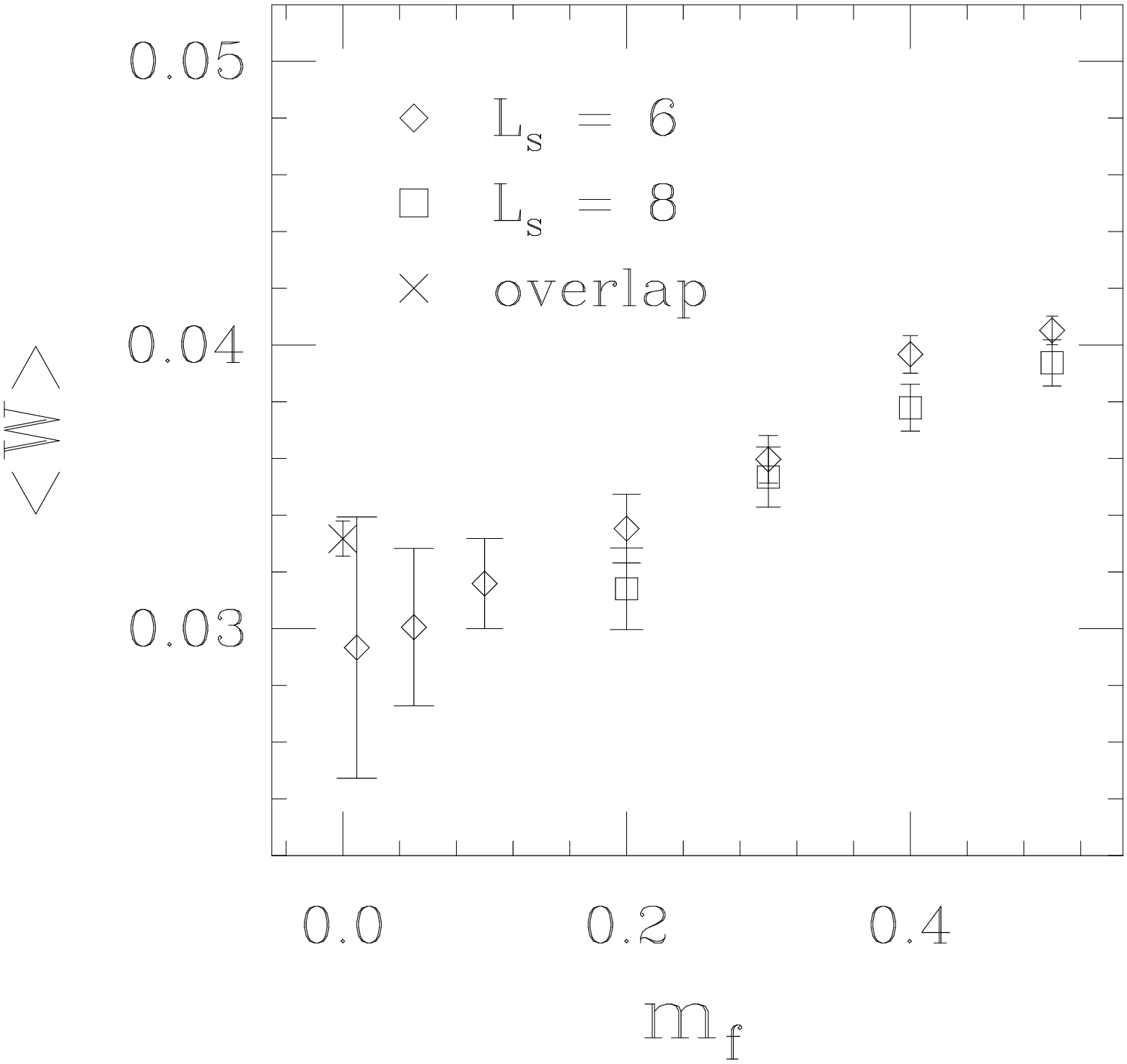}{\four}{\figsize}

\end{document}